\documentclass[prb,twocolumn,showpacs,showkeys,floatfix]{revtex4}

\usepackage{amsmath,amssymb}	
\usepackage{graphicx}
\usepackage{dcolumn}

\begin{document}
\bibliographystyle{apsrev}

\title{Hard x-ray spectroscopy in Na$_x$CoO$_{2}$ and superconducting Na$_x$CoO$_{2}\cdot y\,$H$_2$O:\\ A view on the bulk Co electronic properties}
\author{Ph.~Leininger}
\affiliation{Laboratoire de Chimie Physique--Mati\`{e}re et Rayonnement (UMR~7614), Universit\'e Pierre et Marie Curie, 11 rue Pierre et Marie Curie, 75231 Paris Cedex~05, France}
\author{J.-P. Rueff}
\altaffiliation[Present address: ]{Synchrotron SOLEIL, L'Orme des Merisiers, 
Saint-Aubin, B.P.~48, 91192 Gif-sur-Yvette Cedex, France}
\affiliation{Laboratoire de Chimie Physique--Mati\`{e}re et Rayonnement (UMR~7614), Universit\'e Pierre et Marie Curie, 11 rue Pierre et Marie Curie, 75231 Paris Cedex~05, France}
\author{A. Yaresko}
\affiliation{Max-Planck-Institut f\"{u}r Physik komplexer Systeme, N\"{o}thnitzer Stra{\ss}e~38, 01187~Dresden, Germany}
\author{O. Proux}
\affiliation{European Synchrotron Radiation Facility, 6 rue Jules Horowitz, B.P.~220, 38043 Grenoble Cedex~9, France}
\author{J.-L. Hazemann}
\affiliation{European Synchrotron Radiation Facility, 6 rue Jules Horowitz, B.P.~220, 38043 Grenoble Cedex~9, France}
\author{G. Vank\'o}
\affiliation{European Synchrotron Radiation Facility, 6 rue Jules Horowitz, B.P.~220, 38043 Grenoble Cedex~9, France}
\author{T. Sasaki}
\affiliation{Institute for Materials Research, Tohoku University, Katahira 2-1-1, Aoba-ku, Sendai 980-8577, Japan}
\author{H. Ishii}
\affiliation{National Synchrotron Radiation Research Center, Hsinchu 30076, Taiwan}
\author{J.-M. Mariot}
\affiliation{Laboratoire de Chimie Physique--Mati\`{e}re et Rayonnement (UMR~7614), Universit\'e Pierre et Marie Curie, 11 rue Pierre et Marie Curie, 75231 Paris Cedex~05, France}
\date{\today}

\begin{abstract}
The electronic properties of Co in bulk Na$_{0.7}$CoO$_{2}$ and the superconducting hydrated compound Na$_{0.35}$CoO$_{2}\cdot y\,$H$_{2}$O have been investigated by x-ray absorption spectroscopy (XAS) and resonant inelastic x-ray scattering (RIXS) using hard x-rays. The XAS spectra at the Co $K$-edge were measured in both compounds with two different polarization directions. The changes in the XAS spectra upon hydration and their polarization dependence are well accounted for by linear muffin-tin orbital calculations in the local density approximation. The underlying electronic structure indicates the strong hybridization between the Co $3d$ and O $2p$ states in both compounds, while the electron localization is enhanced in the hydrated compound due to the increase of the Co-Co interplanar distance. The Co $K$ pre-edge further highlights the splitting of the $d$ band as a result of the crystal field effect and demonstrates the Co valency increase when Na$_{0.7}$CoO$_{2}$ is hydrated. The RIXS spectra measured at the Co $K$-edge show an energy loss feature around 10~eV in both compounds in fair agreement with the calculated dynamical structure factor. The RIXS feature is associated to a damped plasmon excitation.
\end{abstract}
\maketitle

\section{Introduction}
\label{sec:Introduction}
The recent discovery of superconductivity in Na$_x$CoO$_2 \cdot y$H$_2$O (Ref.~\onlinecite{Takada2003}) has renewed interest in layered cobaltates. These compounds have been extensively studied in the past for their large thermoelectric power. But the novel hydrated material, the first to show superconductivity, calls for a new look at their electronic properties. The parent compound Na$_x$CoO$_2$ has a hexagonal structure (space group~194; P6$_3$/\textit{mmc}) consisting of alternating Na- and CoO$_2$-layers stacked along the $c$-axis. The superconducting compound is obtained by the hydration of the parent compound whose main structural effect is the dramatic increase (about a factor of two) of the Co-Co inter-planar distance upon intercalation of ice-like sheets. The critical temperature $T_c$ is observed to be weakly dependent on Na doping, with a maximum  at $\approx 5$~K for $x=0.35$.\cite{Schaak2003,Chen2004} 

The strong anisotropy of the resistivity between the in-plane and the out-of-plane directions reported in these compounds indicates quasi two-dimensional electronic properties~\cite{Chou2004,Terasaki1997} and underlines the important role played by the Co-O planes, in particular for superconductivity. The Co-O planes form a spin frustrated triangular lattice of edge sharing CoO$_6$ octahedra. This strongly differs from the high-$T_c$ cuprates, where the Cu-O planes are arranged in a square lattice of corner sharing oxygen octahedra. Furthermore in cobaltates, the Co valency is supposedly mixed between 4+ ($3d^5$) and 3+ ($3d^6$) depending on the Na content $x$, but the exact valent state of Co is ill-defined: Early works have indicated that the average Co valency $\bar{n}$ in the hydrated compound, as estimated from chemical titration, departs from the ($\bar{n}=4-x$) value expected from charge neutrality.\cite{Takada2004,Milne2004} The difference has been attributed to the presence of H$_3$O$^{+}$ ions in the Na layers in the hydrated compound. In contrast, recent nuclear magnetic resonance data have showed that the Co valence state is insensitive to hydration and only depends on the Na content.\cite{Mukhamedshin2005} A statement confirmed lately by powder neutron diffraction.\cite{Viciu2006}

The indeterminate Co valency in these materials reveals the difficulty to properly characterize the Co $d$ states and their hybridization with the neighboring atoms. In the presence of the crystal electric-field in the octahedral environment, the Co $3d$ level is split into a triply degenerated $t_{2g}$ level and a doubly degenerated e$_g$ level. The trigonal distortion of the CoO$_6$ octahedra lifts the degeneracy of the $t_{2g}$ states to give one $a_g$ and two $e_g'$ levels. A critical issue related to the Co properties is the supposed existence of ``hole-pockets'' formed by these $e_g'$ electrons at the Fermi surface. The nesting conditions of these hole-pockets might be responsible for the onset of superconductivity in the hydrated material.\cite{Johannes2004a,Rueff2006} This Fermi surface topology was first proposed by density functional theory calculations in the local density approach (LDA)~\cite{Singh2000,Johannes2004} and later confirmed in presence of dynamical Coulomb correlations.\cite{Ishida2005} But no such Fermi surface topology has been observed in photoemission measurements.\cite{Hasan2004,Yang2004} More recently, a calculation using a multi-orbital Hubbard model in the Hartree--Fock and strong-coupling Gutzwiller approximations showed the absence of hole-pockets in the Fermi surface,\cite{Zhou2005} in better agreement with experiments. 

One reason for the lack of consistency in the characterization of the Co electronic properties lies in the sensitivity of the materials to humidity. Surface-clean samples are difficult to prepare and specific studies of the bulk materials become necessary. In this paper we report on the electronic structure of both the hydrated and non-hydrated compounds, using complementary hard x-ray spectroscopies at the Co $K$-edge, namely x-ray absorption spectroscopy (XAS) and resonant inelastic x-ray scattering (RIXS). These high-energy probes are selective of the cobalt properties and provide bulk sensitive data. The experimental XAS and RIXS spectra will be compared to spectra obtained from first-principles calculations.

The polarization-dependent XAS spectra at the Co $K$-edge ($1s\rightarrow 4p$ transition) are discussed in Sec.~\ref{sec:XAS}. The XAS data are analyzed in the light of simulations based on the electronic density of states (DOS) given by linear muffin-tin orbital (LMTO) calculations. In Sec.~\ref{sec:RIXS}, we address the low-energy excitations observed by RIXS at the Co $K$-edge. The results are tested against calculations of the optical absorption spectrum derived from the computed DOS. Conclusions are drawn in Sec.~\ref{sec:Conclusion}.

\section{C{\lowercase{o}} $K$-edge x-ray absorption}
\label{sec:XAS}
\subsection{Experiments}
\label{subsec:XASExp}
The experiments were performed on single-crystals of Na$_{0.35}$CoO$_{2} \cdot 1.3\,$H$_{2}$O (NCOH) and Na$_{0.7}$CoO$_{2}$ (NCO). The samples were prepared in the form of thin plates ($\approx~0.5$~mm thick, $1.5 \times 0.6\mathrm{~mm}^2$ for NCOH and $1 \times 2\mathrm{~mm}^2$ for NCO) with the $c$-axis normal to the sample surface. The Co $K$-edge XAS  spectra were acquired at the BM30 beamline at the European Synchrotron Radiation Facility (Grenoble), in total fluorescence yield mode. Details of the experimental setup have been described elsewhere.\cite{Proux2005} The incident energy is selected by a double crystal Si(220) monochromator providing a total energy bandwidth of 450~meV. The beam was focused in the horizontal plane by the second monochromator crystal and in the vertical plane thanks to a Rh-coated mirror, located downstream. The spot size was $280 \times 150~\mu\mathrm{m}^2$ at the sample position. The samples were maintained between two kapton sheets. The measurements were carried out on both samples at 20~K using a liquid helium cryostat. The low-temperature mount was found to be efficient to retain NCOH (NCO) hydrated (non-hydrated).\cite{Rueff2006,Chainani2004} The spectra were measured in reflection geometry along two polarization directions by turning the sample holder by $80^{\circ}$ in the horizontal plane, starting from the polarization {\boldmath{$\epsilon$}} quasi perpendicular to  $\mathbf{c}$ (in-plane polarization) to {\boldmath{$\epsilon$}}  parallel to the $\mathbf{c}$ (out-of-plane polarization). In our experimental geometry, the detector was always perpendicular to the incident beam. 

\begin{figure}[b!]
\includegraphics[width=7.5 cm]{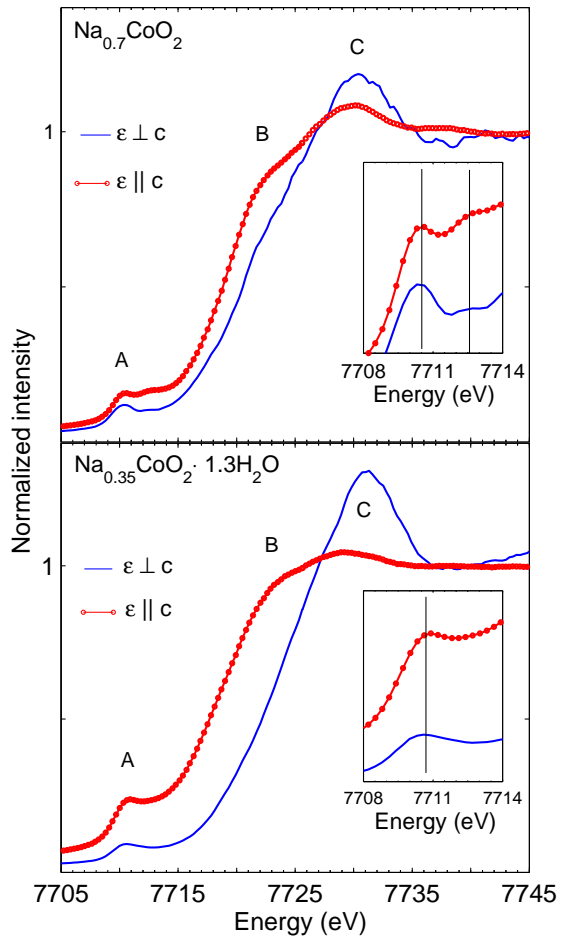}
\caption{(Color online) Co $K$ XAS spectra in Na$_{0.7}$CoO$_2$ (upper panel) and Na$_{0.35}$CoO$_2 \cdot 1.3\,$H$_2$O (lower panel) with {\boldmath{$\epsilon$}} $\perp \mathbf{c}$ (solid lines) and {\boldmath{$\epsilon$}} $\parallel \mathbf{c}$ (circles). The insets show the corresponding enlarged pre-edge region for both compounds. Vertical lines indicate the peak energy positions.}
\label{fig:xas_new} 
\end{figure}
Figure~1 shows the polarization dependence of the Co $K$-edge XAS spectra in NCOH (lower panel) and NCO (upper panel). The spectra are normalized to an edge jump of unity. The spectra exhibit a low-intensity feature (labeled A in Fig.~1) in the pre-edge region at 7710.9~eV (7710.5~eV) in NCOH (NCO) followed by two pronounced peaks (respectively labeled B and C) at 7723.5~eV (7723.5~eV) and 7731.0~eV (7730.4~eV) in the vicinity of the absorption edge. 
When the polarization is turned from in-plane to out-of-plane, we notice a strong decrease in the intensity of the white line C in both samples, whereas peak B increases. The polarization dependence indicates that peak C is mainly associated to $1s\rightarrow 4p_{x,y}$ (in-plane) transitions, while peak B is related to $1s\rightarrow 4p_z$ (out-of-plane) transitions. 

No sign of Co$^{3+}$--Co$^{4+}$ charge disproportionation is observed in the near-edge region. This contrasts with recent Co $2p$ photoemission measurements in NCO and NCOH where well-separated features seemingly related to Co$^{3+}$ and Co$^{4+}$ have been observed.\cite{Chainani2004} Similarly to core-level photoemission, XAS is a fast probe and should be able to distinguish between  mixed valent states, whether static or dynamic order exists. However, the description of the ground state in terms of ionic configurations is not valid here because of covalency effects and multiple charge-transfer states have to be taken into account. In the XAS final state, the different configurations are split in energy because of the core-hole potential, but with additional intermixing. The XAS spectral features are therefore not directly assignable to pure valent states. A similar observation was made at the Co $K$-edge in the mixed-valent Co$^{3+}$/Co$^{4+}$ compounds La$_{1-x}$Sr$_x$CoO$_3$~(Ref.~\onlinecite{Sunstrom1998}).

The pre-edge region, on the contrary, is in principle more sensitive to the Co valent states: in transition metal compounds, this spectral region mainly results from quadrupolar $1s\rightarrow 3d$ transitions, though with an admixture of dipolar character because of the trigonal distortion of the Co octahedral site.\cite{Arrio2000} A closer look shows that the pre-edge consists of two peaks in NCO (see vertical lines in the inset to Fig.~\ref{fig:xas_new}) which we relate to the crystal-field splitting (mainly of octahedral symmetry) of the Co $3d$ states. The low-energy peak corresponds to transitions to the empty $t_{2g}$ states, whereas the peak at higher energy is associated to transitions to the $e_g$ states. This energy splitting ($\approx 2~\mathrm{eV}$ in NCO) is consistent with the values of the crystal-field parameter $10Dq$ used in cluster model simulations of the Co $2p$ photoemission spectra of NCO and NCOH (estimated at 2.5~eV and 4.0~eV respectively for Co$^{3+}$ and Co$^{4+}$ ions).~\cite{Chainani2004} In NCOH, only the lowest $3d$ split state is visible (see vertical line in the inset to Fig.~\ref{fig:xas_new}). We first note that the dipolar-like tail of the white line extends further towards lower energy in NCOH compared to NCO, thus notably masking the high-energy crystal-field related structures. More importantly, the change in the pre-edge region in  NCOH primarily reflects the variation of the Co valency, induced by the different Na doping level. In NCOH, the number of holes in the $t_{2g}$ states (0.65) is about a factor 2 larger than in NCO~(0.3). Assuming that the strength of quadrupolar matrix elements does not depend on the sample, this will lead to enhanced intensity of the lowest energy absorption peak corresponding to transitions to empty $t_{2g}$ states in NCOH compared to NCO, with respect to the $e_{g}$ peak, as observed in the spectra. The spectral changes in the pre-edge region therefore confirm the effects of Na doping on the Co valency. 

In order to analyze further the changes in the XAS spectra as a function of both polarization and Na doping, the Co $K$-edges were simulated from LMTO band structure calculations. This one-electron approach is well suited here because the XAS final states are mainly composed of Co $4p$ band-states. The monoelectronic approach is not expected to reproduce every feature of the absorption spectra but, as explained hereafter, provides valuable information on the Co electronic properties in these materials.

\subsection{Calculations}
\label{XAS:calc}
\subsubsection{Density of states}
\begin{figure}[h!]
\includegraphics[scale=0.29]{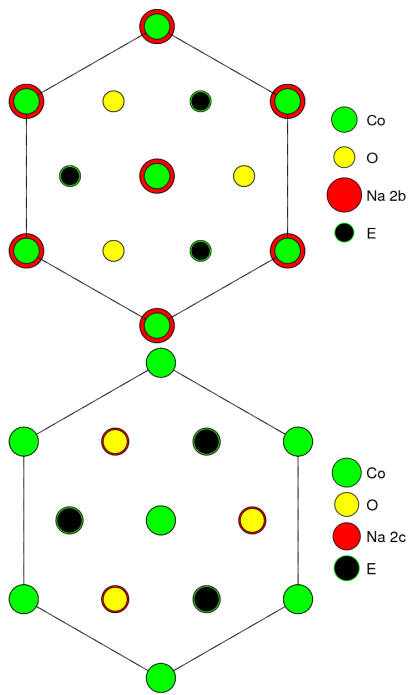}  
\caption{(Color online) Atomic and empty (E) spheres positions in the NCO unit cell with Na occupying site $2b$ (upper graph) and $2c$ (lower graph). All spheres are projected onto the ($\mathbf{a}$, $\mathbf{b}$) plane.}
\label{fig:es} 
\end{figure}

The calculations were performed using an LMTO code~\cite{Perlov} within the local density approximation starting from the known crystal structures for both compounds.\cite{Lynn2003}
We used the von~Barth--Hedin parametrization for the exchange-correlation potential.\cite{Barth1972} Spin-polarization was neglected. One empty sphere per unit cell was used in order to minimize the overlap of the atomic spheres. The positions of the empty and atomic spheres in the unit cell are shown in Fig.~\ref{fig:es} with Na alternatively occupying the $2b$ and $2c$ sites (upper and lower graphs). 
Separate band structure calculations were performed for each compound with Na either in $2b$ or $2c$ sites. Since the exact partitioning of Na between the two sites is unknown, the calculated DOS and absorption spectra were averaged assuming that the $2b$ and $2c$ sites are equally populated. To obtain the right Na stoichiometry (i.e., 0.35 in NCOH and 0.7 in NCO), the Na concentration was adjusted artificially using the virtual crystal approximation (VCA). In this approximation, the Na atoms are replaced by ``virtual atoms'' defined by a pseudopotential taken as the average potential of the real atoms at the same site. In the calculations, this was achieved by allowing for a non-integer Na nuclear charge and, hence, a non-integer number of valence electrons. The VCA approximation was used by Singh \emph{et al.}~\cite{Singh2000} for band structure calculations in NCO and was shown to yield a satisfactory description of the electronic properties. Finally, the hydrated compound was modeled by constructing a unit cell with an elongated $c$-axis (considered to be the main consequence of the hydration) and displaced O atoms, but without water molecules.

\begin{figure}[t!]
\includegraphics[width=7.5 cm]{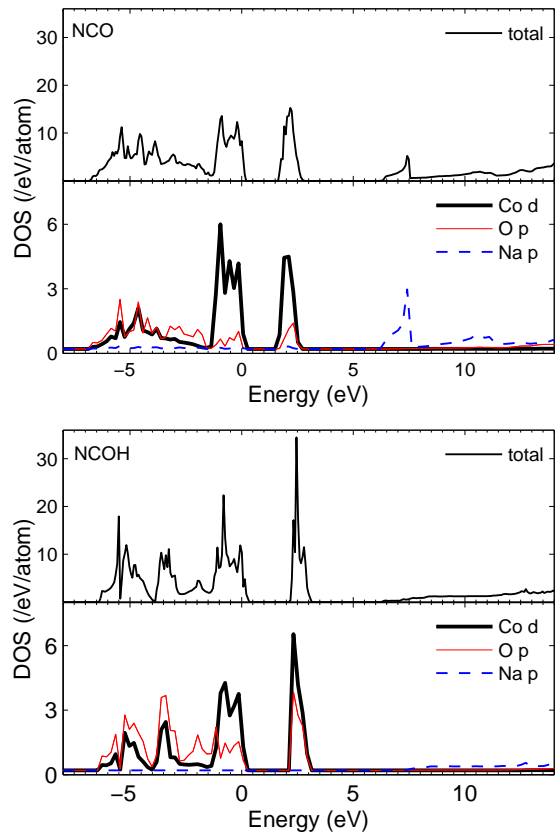}
\caption{(Color online) Calculated total and partial Co $d$, O $p$, and Na $p$ DOS in NCO (upper graph) and NCOH (lower graph).}
\label{fig:dostot} 
\end{figure}

\begin{figure}[t!]
\includegraphics[width=7.5 cm]{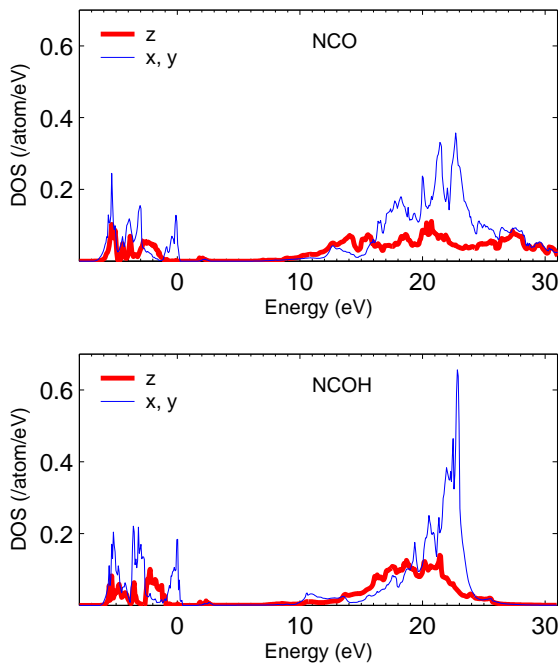}
\caption{(Color online) Partial DOS projected onto Co $p_{x,y}\,(e_u)$ (thin lines) and $p_z\,(a_{2u})$ states (thick lines) in NCO (upper graph) and NCOH (lower graph).}
\label{fig:p_dos} 
\end{figure}

The total and the partial DOS are given in Fig.~\ref{fig:dostot}. We show only the Co $d$, O $p$,  and Na $p$ DOS, as they are the main contributions to the total DOS. They are in good agreement with previous calculations,\cite{Singh2000,Gao2004,Zou2004,Okabe2004} although we concentrate here on a more expanded energy scale compared to the cited works and not to the details of the valence electrons in the region of the Fermi energy ($E_F$). Far below $E_F$ (from $-6.0$~eV to $-1.5$~eV), the DOS has mainly O $p$ character in addition to a weaker contribution of the Co $d$ states. The close connection between the O- and Co-projected DOS indicates a strong $p$-$d$ hybridization of the occupied states. In both compounds the DOS show a non-vanishing intensity at $E_F$, consistent with their metallic character. The split Co-$d$ band, striding the Fermi level, is reminiscent of the crystal-field effect. The splitting amounts to 2.4~eV and 2.9~eV in NCO and NCOH (measured from the position of the center of mass of each $d$ subband). The former value is slightly higher than the 2~eV splitting found experimentally in NCO. Such a discrepancy is common within the LDA, which is known to overestimate hybridization of $d$ orbitals. At higher energy, the DOS mainly consists of Na $p$ states hybridized with Co $p$ (see Fig.~\ref{fig:p_dos}) and O $p$ states. We note that the total DOS in the hydrated compound shows sharper features compared to the non-hydrated one. This could indicate a stronger localization of the electrons in the hydrated compound resulting from a nearly doubling of the $c$ lattice constant.

The partial DOS projected onto the Co $p_{x,y}\,(e_u)$ and $p_z\,(a_{2u})$ states are shown in Fig.~\ref{fig:p_dos}. The empty $p$ states are spread over a large energy range ($\approx 10$--30~eV in NCO; $\approx 10$--25~eV in NCOH). The elongation of the $c$-axis in the hydrated compound likely is at the origin of the bandwidth decrease. This affect mostly the $p_z$ component, as presumed from steric arguments. More unexpectedly, the calculations indicate that the $p_{x,y}$ states are far from being insensitive to the structural change. 
\begin{figure}[t!]
\includegraphics[width=7.5 cm]{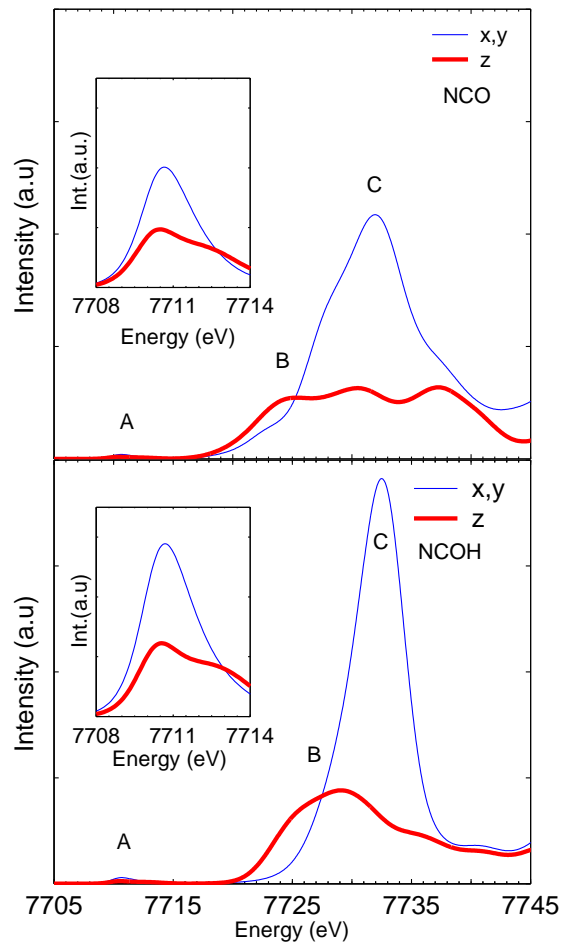}
\caption{(Color online) Calculated Co $K$-edge XAS spectra in NCO (upper graph) and NCOH (lower graph) for the two polarization directions, in-plane (thin lines) and out-of-plane (thick lines). The pre-edge region is enlarged in the insets.}
\label{fig:xas_calc} 
\end{figure}

\subsubsection{Co $K$-edge XAS spectra}
The Co $K$-edge XAS spectra were calculated by multiplying the DOS by the energy dependent dipolar matrix elements. The matrix elements are calculated at each $\mathbf{k}$-point, before integration over the Brillouin zone. Figure~\ref{fig:xas_calc} shows the XAS spectra calculated for the two polarization directions, in-plane (thin lines) and out-of-plane (thick lines) in the hydrated (lower graph) and the non-hydrated (upper graph) compounds. The contribution from the continuum is not included. The spectra are shown as calculated without normalization, the cutoff at $E_F$ being aligned to the experimental value (taken at the flex point of peak~A).

\begin{figure*}[t!]
\includegraphics[width=14.70 cm]{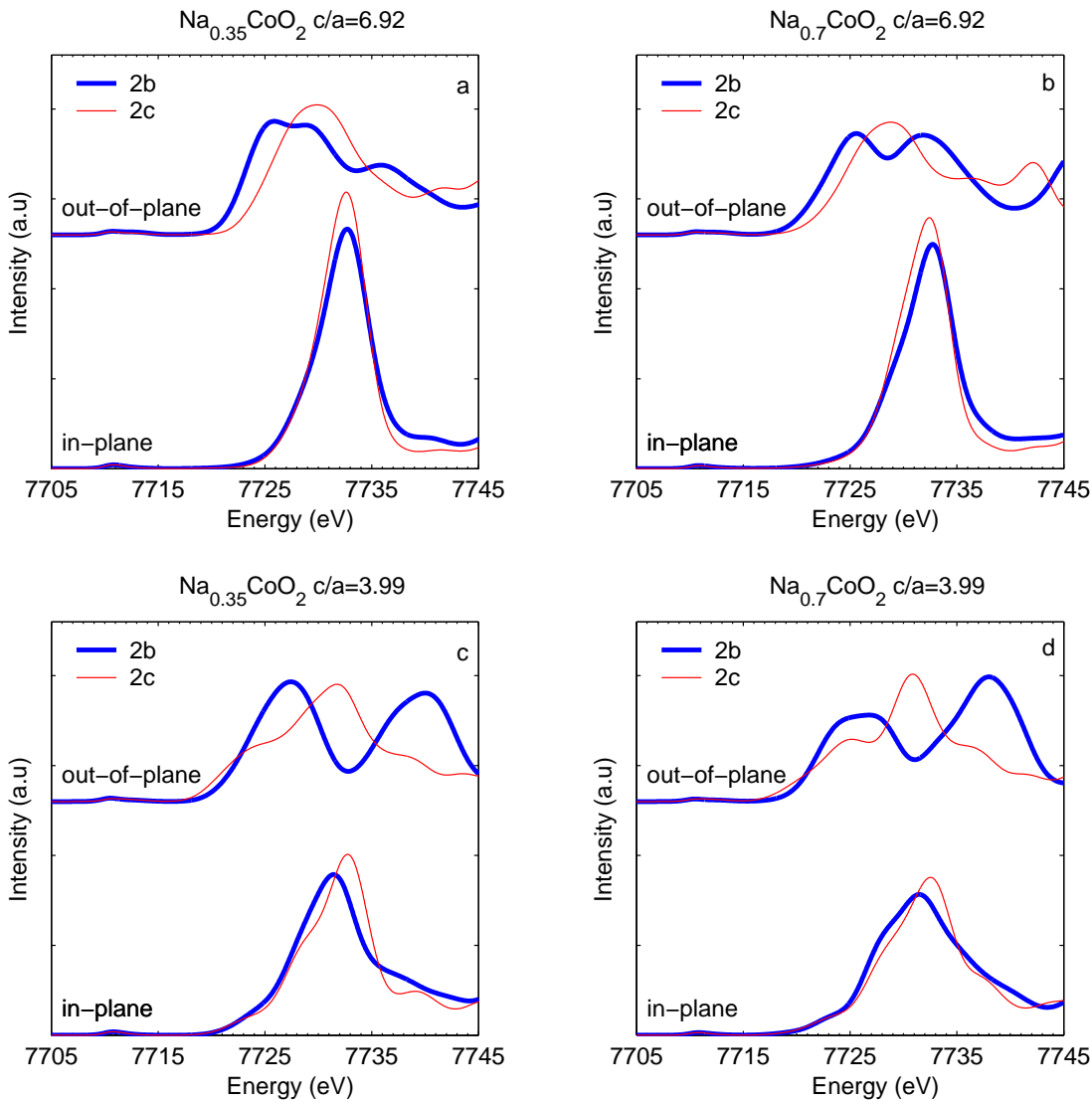}
\caption{(Color online) Calculated Co $K$-edge XAS spectra in Na$_x$CoO$_2$ compound, for two different $c/a$ ratios (3.99 and 6.92, lower and upper graphs, respectively) and for the two different Na sites [$2b$ (thick lines) and $2c$ (thin lines)]. Panels (a) and~(c) [(b) and (d)] represent the calculations with a Na concentration of 0.35 (0.7). In each panel, the lower spectra correspond to the in-plane polarization ($x,y$) direction and the upper spectra to the out-of-plane polarization~($z$).}
\label{fig:calc_4sites} 
\end{figure*}

The calculated spectra reproduce qualitatively the experimental results, though with differences in the peak intensities and positions. In the in-plane calculated spectra, the main line appears at 7732.5~eV in the hydrated compound and 7732.0~eV in the non-hydrated one. These features, which correspond to peak C in the experimental spectra, are shifted to higher energies by $\approx 1.5$~eV. Moreover, a broadening of the white line is observed in the non-hydrated compound compared to the hydrated one, similarly to the experiment. This is likely a  consequence of the more localized character of the electrons in the hydrated compound compared to non-hydrated one, as mentioned above. The observed changes in the ($x,y$) and $z$ DOS are also reflected in the XAS spectra. In the out-of-plane configuration, the spectra are no longer characterized by a white line but present a series of broad features of comparable intensity, specially in NCO. Feature~B, which is located at 7724.0~eV in NCO, is in good agreement with the experimental value with, however, a shift of $\approx$ 0.5~eV to higher energy. In the hydrated compound, the B feature appears broader because of the superimposition, in this energy region, of at least two components as suggested by its asymmetric spectral shape.

We notice in the calculated spectra, a peak located at around 7710.5~eV of sizable intensity (see inset to Fig.~\ref{fig:xas_calc}) which reminds the pre-edge features A observed in the experimental XAS spectra. It cannot be attributed to quadrupolar transitions since only the dipolar contributions are considered in the calculations. It should rather be understood as resulting from the hybridization between the Co $4p$ states with the Co $3d$ states through the interaction with the O $2p$ ligand states. In both NCO and NCOH simulated spectra, we note a $\approx2.4$~eV splitting of the pre-edge in the $z$ direction, reminiscent of the crystal-field effect on the Co$^{3+}$ ion. The splitting is absent in the ($x,y$) component, which could point to a more delocalized character of the in-plane electrons. In the experimental spectra, crystal-field related features were only observed in NCO and in both geometries. Direct comparison with the calculations is difficult however, because quadrupolar and dipolar-assisted transitions are mixed in the experimental spectra. By inspecting the Co $d$ DOS in more detail (not shown here), the calculated pre-edge is seen mainly composed of in plane  Co $p$ ($e_{u}$) states hybridized with Co $d$ ($t_{2g}$) states. Thus, even a relatively small number of holes in the $t_{2g}$ states leads to strong increase in the intensity of the corresponding peak for the in-plane polarization. As a result, the pre-edge intensity in the in-plane XAS spectrum of NCOH will increase compared to NCO.

In the calculated XAS spectra it is important to disentangle the influence of the Na sites from the impact of the Na concentration and the hydration ($c$-axis elongation) effects. With this aim in mind, we show in Fig.~\ref{fig:calc_4sites} the calculated spectra separately for the two Na sites ($2b$ and $2c$) and as a function of the Na concentration (0.35 or 0.7) and the $c/a$ ratio (3.99 and 6.92, for NCO and NCOH respectively). Each spectrum is further projected onto the two polarization directions, i.e., ($x,y$) (lower spectra) and $z$ (upper spectra). For a given $c/a$ ratio, the calculations for two different Na sites ($2b$, $2c$) and two different Na concentrations (0.35, 0.7) are almost identical in the in-plane configuration (especially at $c/a=6.92$). On the contrary, the spectra in the out-of-plane configuration do show differences when substituting a Na on site $2b$ by one on site~$2c$. The largest changes are observed for $c/a=3.99$, when interplane interactions are largest.  Similarly, the variation of the Na concentration mostly affects the out-of-plane spectra obtained with Na occupying the $2b$ site. The highest sensitivity when Na occupies site $2b$ may be related to its location in the unit cell above the Co atoms, while the off-centered Na $2c$ atoms are a much weaker perturbation to the Co $p$ states. 

To summarize, the elongation of the $c$-axis is the main modifying factor of the XAS spectra, thus of the Co electronic properties, while the Na concentration and site-substitution has a less influence. This \textit{a posteriori} justifies our starting hypothesis for the LMTO calculations with Na equally distributed in the $2b$ and $2c$ sites. The out-of plane spectra are observed to be highly sensitive to the structural change as expected from the quasi two-dimensional character of the Co-O network. 

\section{C{\lowercase{o}} $K$-edge resonant inelastic x-ray scattering}
\label{sec:RIXS}
Signatures of Co valence states may be revealed also in the low-energy excitation spectra through intra-band or inter-band excitations around~$E_F$. Electron energy-loss spectroscopy or optical absorption provide means of measuring low-lying excited states.  However, unlike RIXS, they do not benefit from the resonant enhancement and only transition with a very low momentum transfer $q$ can be observed. We analyze here results obtained by RIXS at finite $q$ in NCO and NCOH, in the light of the calculated electronic structure derived in section~\ref{XAS:calc}. 

\begin{figure}[htb]
\includegraphics[width=7 cm]{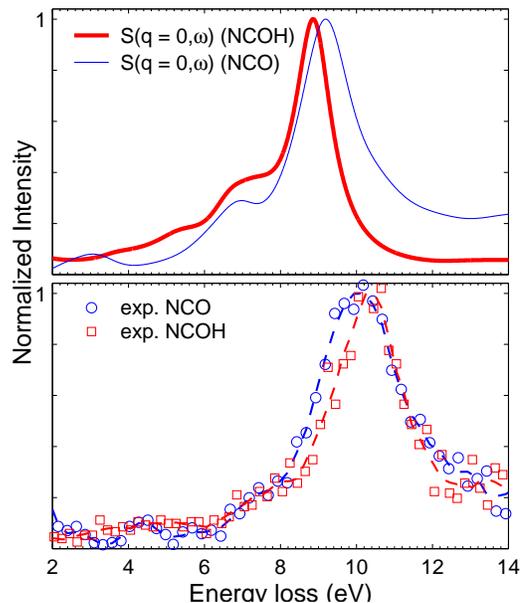}
\caption{(Color online) RIXS spectra at the Co $K$-edge (lower panel) and calculated dynamical structure factor $S(q=0,\omega)$ (upper panel) in NCO and NCOH. For comparison with experiment, the calculated $S$ is convoluted with a 1~eV full width at half-maximum Lorentzian profile. Dotted lines (lower panel) are polynomial smoothed data in NCO and NCOH.}
\label{fig:rixs} 
\end{figure}

The experiment was carried out at beamline ID16 (ESRF) in a previous experimental run using single crystals of NCO and NCOH at the Co K-edge. Details about the measurements and complete data set have been reported in Ref.~\onlinecite{Leininger2006}. Both samples are intrinsically of poor crystalline quality as a result of their synthesis process. To minimize the contribution from the quasi-elastic line, the spectra were measured at a high scattering angle corresponding to a momentum transfer of $q\sim 2$~A$^{-1}$. The main result is summarized in Fig.~\ref{fig:rixs} (bottom panel). The RIXS spectra exhibit a single excitation peaking at 10.0~eV in NCO and at 10.4~eV in NCOH. Note that the spectra of NCOH and NCO were measured with different energy resolutions, respectively 0.4~eV and 1~eV. The inelastic feature resonates in the vicinity of the white line energy (7729~eV). The high energy of this excitation precludes it originating from crystal-field-like excitations. To help the identification of this excitation, we plot in Fig.~\ref{fig:rixs} (upper panel) the calculated dynamical structure factor $S(q=0,\omega)$ in NCO and NCOH. This quantity was derived from the computation of the dielectric response function $\varepsilon$ obtained from the DOS, and according to the relationship $S \propto \mathrm{Im}(-1/{\varepsilon})$. Spectral broadening due to the finite resolution and lifetime effect were taken into account by a convolution of the calculated spectrum with a 1~eV full width at half-maximum Lorentzian profile. The calculation shows an intense peak at $\approx 9.2$~eV in NCO (8.9 eV in NCOH), in good agreement with the RIXS feature, although the latter is shifted to higher energy. The shoulder on the low energy side (clearly seen in the better resolved spectrum in NCOH) is well reproduced as well. The shift between the experiment ($q\neq 0$) and the calculation ($q=0$) may be related to the dispersion of the inelastic features at finite $q$. In the experimental spectra, we observe an apparent displacement of the main peak by 0.5~eV towards high energy (between NCO and NCOH) which contrasts with the calculations. Although the variation of the Co valence with doping could reasonably explain the energy shift \footnote{The plasmon frequency $\omega_p$ for free electrons reads in its derivative form : $\Delta\omega_p/\omega_p=\Delta n/(2n)$. Using $\bar{n}=3.5$ and $\Delta n=0.35$, as expected from the variation of the Co valency between NCO and NCOH, the calculated shift of the plasmon peak is $\Delta\omega_p\approx +\,0.5$~eV.}, we cannot totally exclude resolution effects.  

The overall agreement of the computed dynamical structure factor with the experiment is at first unexpected, considering that the calculated spectrum is derived from a purely non-resonant scattering process, opposite to the resonant cross-section in RIXS. 
Recently, Grenier \emph{et al.}~\cite{Grenier2005} reported similar results in a RIXS experiment performed on manganites at the Mn $K$-edge. They observed an energy-loss feature at 12~eV in the RIXS spectra that they attribute to a plasmon excitation reminiscent of the non-resonant dynamical structure factor. Non-resonant contributions to RIXS can be understood as arising from a third-order process~\cite{Abbamonte1999,Brink2005} which involves shake-up phenomena in the RIXS intermediate states. In this picture, the RIXS cross-section would be proportional to the dynamical structure factor multiplied by a resonant denominator. 

In our case, the momentum transfer is about one order of magnitude higher than that of Ref.~\onlinecite{Grenier2005}. At high $q$, plasmon excitations are severely damped by creation of electron-hole pairs, but the critical wave vector above which the plasmon looses its definition is unknow in the present case without a detailed knowledge of the dispersive behavior. In a systematic IXS study, Gurtubay \emph{et al.}~\cite{Gurtubay2005} lately reported plasmon-like excitations at similar momentum transfer in elemental transition metals, although significantly broader in energy compared to low $q$ measurements. Thus, the main feature in our RIXS spectra is understood as a damped plasmon involving Co $p$-electrons while the fine structure (e.g. low energy shoulder) is attributed to a background of excitations due to inter-band transitions.  

\section{Conclusions}
\label{sec:Conclusion}
We have investigated the Co electronic properties in Na$_{0.7}$CoO$_2$ and Na$_{0.35}$CoO$_2\cdot 1.3\,$H$_2$O using hard x-ray spectroscopic probes. The experimental results are compared to first-principles calculations of the Co $K$-edge XAS spectra within the LMTO approach. The underlying DOS show the strong hybridization of the Co $4p$ band with O $2p$ and Co $3d$ states. This particularly concerns the spectral region close to the Fermi energy. A stronger electron localization is revealed in the hydrated compound with respect to the non-hydrated one, consecutively to the elongation of Co-Co inter-planar distance. The pre-edge region is found mostly sensitive to the Co-valence change, while no clear dependence on the Na content is observed in the near-edge part of the spectra. Detailed calculations indicate that the XAS spectra are more affected by the expansion of the $c$-axis when NCO is hydrated than by the change of the Na doping level. Low-energy excitations observed by RIXS at the Co $K$-edge could be assigned to a damped plasmon mode, presumably involving the Co valence electrons.


\end{document}